\documentclass[aps,pre,reprint,groupedaddress]{revtex4-1}

\newcommand{\lb}{\overline{\ell}}

\newcommand{\bea}{\begin{eqnarray}}
\newcommand{\eea}{\end{eqnarray}}

\usepackage{graphicx}% Include figure files
\usepackage{dcolumn}% Align table columns on decimal point
\usepackage{bm}% bold math

\begin{document}

\title{On the structural properties of small-world networks with finite range of shortcut links}

\author{Tao Jia}
\email{tjia@vt.edu}
\author{Rahul V. Kulkarni}
\email{kulkarni@pooh.phys.vt.edu} 
\affiliation{Department of Physics, \\
Virginia Polytechnic Institute and State University, \\
Blacksburg, VA 24061}

\date{\today}% It is always \today, today,
             %  but any date may be explicitly specified

\begin{abstract}
We explore a new variant of Small-World Networks (SWNs), in which an additional parameter ($r$) sets the length scale 
over which shortcuts are uniformly distributed. When $r=0$ we have an ordered network, whereas $r=1$ corresponds to the 
original SWN model. These short-range SWNs have a similar degree distribution and scaling properties as the original SWN model. 
We observe the small-world phenomenon for $r \ll 1$ indicating that global shortcuts are not necessary for the small-world effect. 
For short-range SWNs, the average path length changes nonmonotonically with system size, whereas for the original SWN model it 
increases monotonically. We propose an expression for the average path length for short-range SWNs based on numerical simulations 
and analytical approximations.

\end{abstract}

% PACS, the Physics and Astronomy
                             % Classification Scheme.
%\keywords{Suggested keywords}%Use showkeys class option if keyword
                              %display desired

\maketitle

\section{Introduction}

The structural and dynamic properties of small-world networks (SWNs) \cite{watts98} 
have been studied intensively with applications to diverse fields such as sociology,
 biological networks, neural networks and so on \cite{albert02,barahona02,Humphries08, sen03, Mason07, Boccaletti06, bassett06,smit08,Atilgan04,Bartoli07}. 
In the original model, the SWN is 
constructed by adding shortcuts on top of a connected lattice ring (ordered lattice).
The added shortcut  connects a pair of nodes picked randomly. This represents the 
case such that each node in the network has equal probability to make random 
connections to all others regardless of their separation for the  ordered lattice
(also denoted as Euclidean distance). Under this construction, the small-world 
effect appears in SWN with the mean separation between any pairs of nodes scaling 
logarithmically with the network size.

Although there has been intensive work on the structural properties of SWNs, 
the relationship between the maximum length of shortcut and the small-world effect 
has not yet been investigated, to the best of our knowledge. In the original 
SWN model, the maximum shortcut length is half of the network size. On the other
hand, if we limit the maximum shortcut length to be some small value, e.g. 
corresponding to the case that the shortcut can only connect the nearest neighbors,
the network will be identical to a  regular lattice. These limits indicate that by 
tunning the maximum range of shortcut lengths, we can change the network form a 
standard SWN to a regular network. An important question that arises is: what are 
the constraints on the maximum shortcut length in order 
to give rise to the small-world effect?

Besides the motivation stated above, there are other reasons to examine more closely
the role of  shortcut range in SWNs. In the standard SWN model, shortcuts are {\it global} with {\it uniform} length distribution, 
which assumes that the shortcut connection is completely random. However, this may not be an accurate representation for
real-world networks. 
For example, in social networks it is reasonable to assume that the probability to have a connection is higher for 
two individuals living close by as 
compared to two individuals living far away. One simple way to address this issue is to modify the property of the shortcuts 
in SWN. In some previous studies, a variant of SWN was proposed with {\it global} shortcuts but power law distribution of the 
shortcut length \cite{jespersen00,moukarzel02,sen01}. This represents the case that the further apart the 
nodes are on the ordered lattice, the harder it is to have a shortcut between them.

In this work, we study another variant of SWN such that the shortcut length has a {\it uniform} length distribution but with 
limited range (maximum length). We call this model the short-ranged small-world network (SRSWN). In SRSWNs, we introduce a 
new parameter $r$ which represents the maximum proportion of the network that a shortcut can reach. This corresponds to the 
case that one has equal probability to build random connection only within a community. Moreover, as the $r$ value varies between 
0 and 1, the SRSWN converts from a regular lattice to a standard SWN. The investigation of SRSWN will also provide insight into
how the small-world effect depends on the shortcut length. In the following, we will explore the structural properties of 
SRSWN with both numerical and analytical approaches.

\section{Model and Results}

The construction of SRSWN is similar to that of SWNs. We first start with a regular connected lattice with $L=2N$ nodes. Shortcuts are added with probability $p$, with totally $x=Lp$ added. Two types of distance are introduced: the ``Euclidean distance'' is defined as the distance of two nodes along the backbone and the ``minimal distance'' denotes the shortest distance after taking shortcuts. An outgoing shortcut will randomly connect a node within Euclidean distance $rN$, {\it i.e.} a node randomly chosen from $rL$ nearest neighbors. Note that this construction yields the same degree distribution as that of SWN (see Appendix). In this work, we concentrate on the case of small $p$($p<0.01$) values. For each realization of network, we pick a random initial position and apply the breadth-first search algorithm to find the quantities analyzed. The  simulation data presented are averaged over at least 
10000 network realizations.

\subsection{Mean path length $\lb$ and $\lb(n)$}

\begin{figure}[tb]
\begin{center}
\resizebox{7.5cm}{!}{\includegraphics[trim = 15 15 30 30, clip]{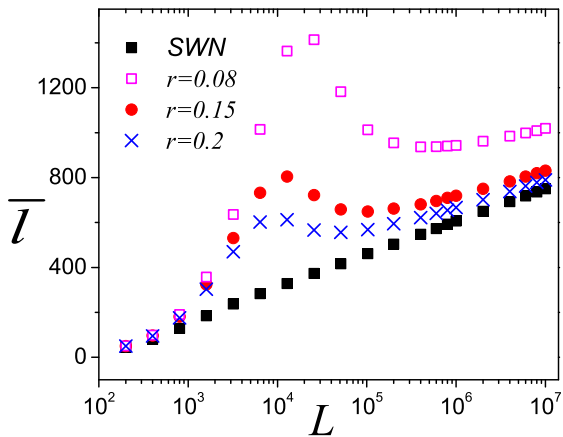}}
\caption{plots of numerical results of $\lb$ vs $L$ with $p=0.004$ and various $r$ values. As shown, $\lb$ changes nonmonotocially in SRSWN while it changes monotocially in SWN. When $L$ is large enough, $\lb$ varies logarithmically with $L$, indicating the appearance of small-world property. \label{fig:l_bar_L}}
\end{center}
\end{figure}\noindent

To check if SRSWN displays small-world phenomenon, we first check the relationship between mean path length $\lb$ (defined as the average distance along the shortest paths for all possible pairs of network nodes) and network size $L$ by simulations. 
The plot of $\lb$ vs $L$ in SRSWN and SWN is shown in Fig[\ref{fig:l_bar_L}] which displays some interesting features: 
(i) Initially $\lb$ for the SRSWN increases almost linearly with $L$, which is similar to a regular network.
(ii) For large values of $L$, $\lb$ for SRSWN has similar behavior as that of SWN, {\it i.e.} $\lb$ scales logarithmically with $L$. 
(iii) In the intermediate region, $\lb$ first reaches a maximum value and then {\em decreases} as $L$ increases. 
This nonmonotonic behavior points to the surprising property that nodes in the network come closer to each other (on average) 
as the network size increases. This feature is not observed in either regular lattice or SWN, as $\lb$ in both cases 
increases monotonically with $L$. Furthermore, point (ii) noted above indicates that SRSWN does show small-world properties as 
long as the network size is large enough.

To gain a further understanding of the numerical observations noted above, we first study $\lb(n)$, which is defined as the minimal path length between two nodes given that the Euclidean separation is $n$. Alternatively, $\lb(n)$ can be considered as the number of time steps a walker needs to take (via the shortest path) from origin to a node at Euclidean distance $n$ away. The relation between $\lb(n)$ and $\lb$ is:
\begin{equation}
\lb=\frac{1}{N} \sum_{n=1}^{n=N}{\lb(n)}. \label{eq:lb_lbn}
\end{equation}

In the regular network, the walker can only walk through the arc of the circle such that $\lb(n) = n$. In SWN, because the shortcut range is global, taking a shortcut will completely randomize the position of the walker. As the shortest path usually contains at least one shortcut and the shortcut can land on any node with same probability, it seems plausible that there will be a large number of nodes having the same probability to be visited in a given $n$ steps. In previous work, it has been shown for large $L$ that $\lb(n) \sim n$ when 
$n < 1/p$ and it quickly saturates to $\lb$ as $n$ increases \cite{Moukarzel99,kulkarni00}. 
This indicated that the majority of the nodes can be reached within the same time steps on average, regardless of their
original positions.

In SRSWN, however, $\lb(n)$ will behave in a different manner from either the regular network or the SWN. First, a shortcut in SRSWN will bring the walker closer to nodes far away and save some time steps, indicating that $\lb(n)$ will be less than $n$ for most of 
the nodes. On another hand, due to the limited range of the shortcut, it cannot take the walker to an arbitrary node as in the SWN, 
indicating that $\lb(n)$ will depend on $n$. Let's examine the case that $r$ is small and focus on the behavior of $\lb(n)$ for 
nodes away from origin ($n=0$) and the extremal node ($n=N$). 
In particular, let us consider three nodes with Euclidean distance $n_1$, $n_2$ and $n_3$ such that 
$n_2 = n_1 + rN$ and $n_3 = n_2 + rN$. Because a shortcut has the maximum length $rN$, node $n_2$ cannot be visited before 
node $n_1$. Clearly we have that $\lb(n_2) > \lb(n_1)$ and $\lb(n_3) > \lb(n_2)$. Denote $\delta \lb_1 = \lb(n_2) - \lb(n_1)$ 
and $\delta \lb_2 = \lb(n_3) - \lb(n_2)$ as the extra time steps the walker takes in going from node $n_1$ to $n_2$ and 
$n_2$ to $n_3$, respectively. Because the network configuration for the interval $[n_1,n_2]$ is similar to that for $[n_2,n_3]$, it 
is reasonable to assume that the walker will, on average, take the same number of time steps in traveling between these two regions, {\it i.e.} $\delta \lb_1 = \delta \lb_2$. Given that the Euclidean separations for the pair of nodes $n_1$,$n_2$ and $n_2$,$n_3$ is the same ($rN$), we expect a linear relationship between $\lb(n)$ and $n$ for $0 \ll n \ll N$.

\par

\begin{figure}[tb]
\begin{center}
\resizebox{7.5cm}{!}{\includegraphics[trim = 15 15 30 25, clip]{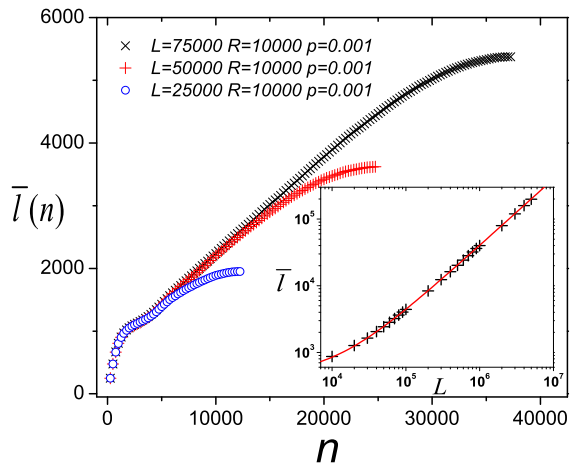}}
\caption{In the main figure, we show numerically computed $\lb(n)$ with different network size $L$ but same $R$ and $p$ values. Apart from the nonlinear behavior near $n \sim 0$ and $n \sim N$, $\lb(n)$ increases linearly with $n$ in the middle region, presumably corresponds to the interval $n \in [rN,(1-r)N]$. The slop of the linear part in the three cases are the same, indicating that the linear increase rate of $\lb(n)$ depends on $R$ and $p$. In the inset, we show the linear relationship between $\lb$ and $L$ with the same choice of $p$ and $R$ values. The points can be well fitted by the line $y=450+0.04x$.\label{fig:l_bar_n}}
\end{center}
\end{figure}

The simulation results for $\lb(n)$ shown in Fig[\ref{fig:l_bar_n}] confirms the expectations from the rough argument given above. 
As displayed, $\lb(n)$ has a part that increases linearly with $n$, which corresponds to the interval $n \in [rN,(1-r)N]$. Furthermore, 
we notice that for networks with different size but same parameter $p$ and $R=rL$, $\lb(n)$ increases with $n$ at the same rate. 
This is because the time steps the walker takes to go through the interval $[n_1,n_2]$ and $[n_2,n_3]$ only depends on the chance of 
encountering a shortcut (respectively $p$) and the range of the shortcut (respectively $rN$ or $R$). As the result, the rate 
corresponding to the linear increase of $\lb(n)$ depends on $p$ and $R=rL$ only.

\noindent

In the limit that $R \ll L$ (equivalently $r \ll 1$) such that the nonlinear behavior of $\lb(n)$ near the beginning and the end can be ignored, we can approximate the linear relationship between $\lb(n)$ and $n$ for all $n$'s from 1 to $N$. Under such an approximation, $\lb$ derived based on Eq.(\ref{eq:lb_lbn}) will be a linear function of $L$ for given $R$ and $p$ values. It is noteworthy that while 
this approximation is for $R \ll L$, we found from simulations (as shown in the inset of Fig[\ref{fig:l_bar_n}]) that the 
linear relationship holds even for $R$ values close to $L$. Given such linear relationship and considering that the minimum 
value of $L$ for given $R$ is $L=R$, we can express the $\lb$ of SRWN with parameter $L$, $p$ and $r$ as
\begin{equation}
\lb(L, p, r)=(L-R)K(R,p)+\lb_{SWN}(R,p), \label{l_bar_eq}
\end{equation}
where $K$ depends on the linear increase rate of $\lb(n)$ that is a function of $R=rL$ and $p$. 
$\lb_{SWN}(R,p)$ denotes the average distance of SWN with size $R$ and shortcut probability $p$, since the network that 
corresponds to $L=R$ is the standard SWN.

\subsection{Scaling properties}

Eq.(\ref{l_bar_eq}) provides a scaling form of $\lb$ in SRSWN. On the other hand, the scaling properties of SWN have been derived from renormalization group transformations \cite{newman99,newman99_2} and this approach can also be generalized for SRSWN. The 
renormalization group transformation consists of conserving the number of the shortcuts while varying network size $L$ and 
shortcut probability $p$. As the parameter $r$ is the length scale of the shortcut range, $\lb$ in SRSWN will continue to be an homogeneous function with $r$ as another variable:
\begin{equation}
\lb(\frac{L}{\lambda},\frac{1}{p\lambda}, r)=\frac{1}{\lambda}\lb(L,\frac{1}{p},r), \label{eq:rg_scale_1}
\end{equation}
which gives another scaling form of $\lb$ as:
\begin{equation}
\lb(L,p,r)=Lg(x,r),\label{eq:rg_scale_2}
\end{equation}
where $g$ is a scaling function and $x=Lp$ is the total number of the shortcuts. The scaling relation in Eq.(\ref{eq:rg_scale_1}) 
has been confirmed by our simulations.

Combining Eq.(\ref{eq:rg_scale_2}) and Eq.(\ref{l_bar_eq}), we have:
\begin{equation}
{\lb \over L} = g(x,r) = (1-r)K(R,p) + \frac{\lb_{SWN}(R,p)}{L}. \label{l_bar_eq2}
\end{equation}
Furthermore, based on the scaling property of SWN \cite{newman99,newman00}, we have
\begin{equation}
{\lb_{SWN}(R,p) \over R} = f(Rp) = f(rx)\label{eq:lb_SWN},
\end{equation}
By combining Eq.(\ref{l_bar_eq2}) and Eq.(\ref{eq:lb_SWN}), we can express the function $g(x,r)$ as
\begin{equation}
g(x,r)=(1-r)K(R,p)+f(rx)r,
\end{equation}
which indicates that $K(R,p)$ is also a function of $x$ and $r$.
By keeping $r$ fixed but choosing different $p$ and $L$ value as $p'=p/\lambda$, $L'=L\lambda$, we keep $x$ and $r$ the same. 
This leads to the following relationship
\begin{equation}
K(rL,p) = K(x,r)= K(\lambda rL, \frac{p}{\lambda}). \label{l_bar_scale}
\end{equation}
The only way to have Eq.(\ref{l_bar_scale}) hold is that $K$ is a function of $rLp$ only.
Eventually, we have the simplified scaling form of $\lb$ from Eq.(\ref{l_bar_eq2}) as:
\begin{equation}
\lb = (L-R)K(y) + Rf(y), \label{eq:l_bar_scale}
\end{equation}
with two scaling functions $K$ and $f$. The variables are $y=rx=rLp$ and $R=rL$.

\subsection{Numerical analysis}

\begin{figure}[tb]
\begin{center}
\resizebox{7.5cm}{!}{\includegraphics[trim = 15 15 30 25, clip]{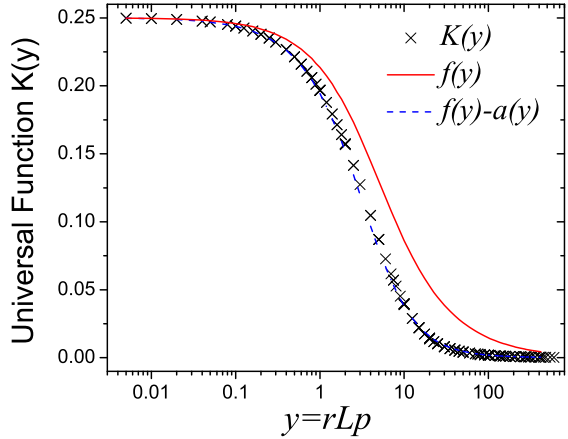}}
\caption{Plots of function $K(y)$ and $f(y)$ based on computational simulations. The dashed line corresponds to the expression $f(y)-\alpha(y)$ found in Eq.(\ref{eq:a(y)_2}) that fits $K(y)$ well. \label{fig:Ky}}
\end{center}
\end{figure}\noindent

\par

The functional form of $f(y)$ in Eq.(\ref{eq:l_bar_scale}) has been studied previously \cite{newman00,newman99_2,almaas02} via both 
numerical and analytical approaches. Particularly when $y \gg 1$, we have 
\begin{equation}
f(y) = log(2y)/(4y) + 0.13/y+O(y), \label{eq:fy}.
\end{equation}
The function $K(y)$ can be determined numerically. To explore the functional form of $K(y)$ and check the scaling property proposed 
in Eq.(\ref{eq:l_bar_scale}), we first run simulations for networks with different size $L_1$ and $L_2$ but same shortcut range $R$ 
and randomness $p$. The mean path length is $\lb_1$ and $\lb_2$, respectively. From on Eq.(\ref{eq:l_bar_scale}), $K$ can be found as $(\lb_1 - \lb_2)/(L_1 - L_2)$. The above process is repeated with various choices of $L$, $R$ and $p$ and each time we have a value of $K$. We then plot the $K$ values found numerically as the function of $y=rLp$. If the scaling form holds, the plot of $K(y)$ will collapse to a single curve, otherwise, the points will be scattered.

\begin{figure}[tb]
\begin{center}
\resizebox{7.5cm}{!}{\includegraphics[trim = 15 15 30 25, clip]{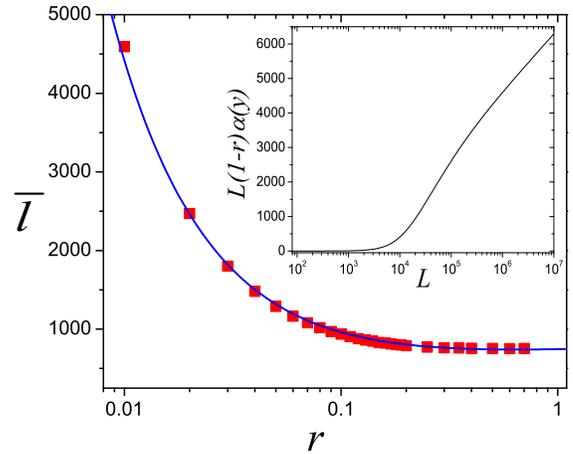}}
\caption{Plots of numerical results of $\lb$ and Eq.(\ref{eq:l_bar_2}) for $L=10,000,000$, $p=0.004$ and various values for $r>0.01$. The choice of the $L,p,r$ values guarantees the satisfaction of $y \gg 1$. The blocks are the numerical results and the line corresponds to Eq.(\ref{eq:l_bar_2}). In the inset, we show the how term $L(1-r)\alpha(y)$ varies with network size $L$ given $r=0.08$ and $p=0.004$. \label{fig:l_bar_r}}
\end{center}
\end{figure}

In Fig[\ref{fig:Ky}], we show the $K(y)$ curve obtained as discussed above. As it displays a single curve as function of $y$, 
the scaling form in Eq.(\ref{eq:l_bar_scale}) is confirmed. To step further, we compared the function $K(y)$ with $f(y)$ and found
that $K(y)$ is very close to $f(y)$ when $y \ll 1$. When $y \gg 1$, $K(y)$ can be well fitted by the function 
$K(y)=\frac{0.16}{y}+\frac{2.91}{y^2}$, which coincides with the second and other higher order terms in Eq.(\ref{eq:fy}). 
Based on these observations, we propose the expression of $\lb$ for large $y$ values as:
\begin{eqnarray}
\lb &=& (L-R)\frac{2.91+0.16y}{y^2}+R(\frac{log(2y)}{4y}+\frac{0.13}{y})\nonumber \\
&\approx& \frac{log(2rLp)}{4p}+\frac{0.16}{rp}, \label{eq:l_bar_2}
\end{eqnarray}
which fits the simulation results quite well as shown in Fig[\ref{fig:l_bar_r}].

\noindent

In Eq.(\ref{eq:l_bar_2}), $\lb$ changes logarithmically with $L$, which confirms our previous observation of small-world effect for
 large $L$ values. More significantly, it indicates that the small-world effect appears when $y=rLp \gg 1$. It further indicates that 
global shortcut range is not necessary for small-world properties: even if the shortcut covers only a small fraction of the network, 
the small-world phenomenon will still occur as long as the network size is large enough.

Eq.(\ref{eq:l_bar_2}) gives the expression of $\lb$ in one limiting case. To find the full expression of $\lb$, we need the numerical fit of $K(y)$ for all $y$ values. Based on the form of $K(y)$ we found in the two limiting case, we related $K(y)$ and $f(y)$ in a way that
\begin{equation}
K(y)=f(y)-\alpha(y),\label{eq:a(y)_1}.
\end{equation}
$\alpha(y)$ is a positive function that approaches to 0 when $y \rightarrow 0$ and goes to $log(2y)/(4y)$ when $y \gg 1$. Using numerical fitting, we propose one possible form of $\alpha(y)$ as
\begin{equation}
\alpha(y)=\frac{log(1+\frac{2y^2}{24+y})}{4y}, \label{eq:a(y)_2}
\end{equation}
which is displayed in Fig.[\ref{fig:Ky}].

The $\lb$ in Eq.(\ref{eq:l_bar_scale}) can then be expressed as a combination of $f(y)$ and $\alpha(y)$ as:
\begin{equation}
\lb = Lf(y) - L (1-r) \alpha(y).
\end{equation}
The first term is equivalently the mean path length of a SWN with rescaled randomness $p'=rp$, which always increases with $L$. 
The second term will be close to 0 when $L$ is small and then increase rapidly when $L$ is large enough, as shown in the inset 
of Fig.(\ref{fig:l_bar_r}). This explains the nonmonotonic behavior of $\lb$ in SRSWN shown in Fig[\ref{fig:l_bar_n}]. 
When $L$ is small, the first term will dominate and $\lb$ will increase with $L$. 
After $L$ reaches a critical value, the term $L\alpha(y)$ will become non-negligible which can take the value of $\lb$ down. 
This corresponds to the part where $\lb$ drops as $L$ increases. Finally, when $L$ is large such that $rLp \gg 1$, 
$lb$ follows Eq.(\ref{eq:l_bar_2}) and increase logarithmically with $L$, similar to that of SWN. The difference of $\lb$ between SRSWN and SWN derives from the term 
$\frac{0.64(1-r)+log(r)}{4pr}$, which yields large values for small $r$ and $p$. This is why $\lb$ of SRSWN is larger than that of SWN 
even when $L$ is large. (Note that the term $\frac{0.64(1-r)+log(r)}{4pr}$ can be negative for $r>0.5$, which is mostly
due to the error in numerical fit. But its absolute value is very small and we consider the error as acceptable.)

\par

\section{Summary}

In conclusion, we have studied a new variant of SWN model called SRSWN, where the shortcut length is constrained within $rN$. 
The SRSWN shows the same degree distribution (see Appendix) and similar scaling properties as SWN. However, it also exhibits 
distinctive structural features which are not seen in SWNs. The most striking of these is
 the nonmonotonic behaviour of the avergae path length: for a range of system sizes, 
as the number of nodes increases the mean path length decreases, i.e. the network appears to shrink as the number of nodes increases.
It would be of interest to see if this behavior is 
also reflected in more realistic networks which 
are expected to have  a finite range of shortcut links. Our results also demonstrate that a global range for the shortcuts 
is not necessary to obtain the small-world effect. For SRSWNs, we derive analytical expressions for quantities of interest 
such as the mean path length which are in excellent agreement with simulations. The results obtained thus suggest that 
similar approaches could be used to accurately characterize 
structural properties of diverse networks which build on the original SWN model to 
include more realistic generalizations.

The authors acknowledge funding support from  NSF (PHY-0957430) and from ICTAS, Virginia Tech. 

\section{Appendix}

The distribution $p(l)$ is defined as the probability that a node has $l$ shortcuts attached to it. In SRSWN, $p(l)$ amounts to
\begin{eqnarray}
p(l) &=& p {rL-1\choose l-1}(1-\frac{p}{rL-1})^{rL-l-2}(\frac{p}{rL-1})^{l-1} \nonumber \\
&+& (1-p){rL-1\choose l}(1-\frac{p}{rL-1})^{rL-l-1}(\frac{p}{rL-1})^l.  \label{eq:degree_distrib_1}
\end{eqnarray}
The first term denotes the case such that there is one outgoing shortcut and $l-1$ incoming shortcuts, while the second term is the 
case that all $l$ shortcuts are incoming ones.

In this paper, we focus on the case when $R=rL$ value large. In this case, Eq.(\ref{eq:degree_distrib_1}) converges to a 
combination of two Possion distributions as
\begin{equation}
p(l)=p{p^{l-1} \over (l-1)!}e^{-p}+(1-p){p^l \over l!}e^{-p}.\label{eq:degree_distrib_2}
\end{equation}
which is the same as that of SWN.

\bibliographystyle{apsrev}
\bibliography{network}

\begin{thebibliography}{20}
\expandafter\ifx\csname natexlab\endcsname\relax\def\natexlab#1{#1}\fi
\expandafter\ifx\csname bibnamefont\endcsname\relax
  \def\bibnamefont#1{#1}\fi
\expandafter\ifx\csname bibfnamefont\endcsname\relax
  \def\bibfnamefont#1{#1}\fi
\expandafter\ifx\csname citenamefont\endcsname\relax
  \def\citenamefont#1{#1}\fi
\expandafter\ifx\csname url\endcsname\relax
  \def\url#1{\texttt{#1}}\fi
\expandafter\ifx\csname urlprefix\endcsname\relax\def\urlprefix{URL }\fi
\providecommand{\bibinfo}[2]{#2}
\providecommand{\eprint}[2][]{\url{#2}}

\bibitem[{\citenamefont{Watts and Strogatz}({1998})}]{watts98}
\bibinfo{author}{\bibfnamefont{D.}~\bibnamefont{Watts}} \bibnamefont{and}
  \bibinfo{author}{\bibfnamefont{S.}~\bibnamefont{Strogatz}},
  \bibinfo{journal}{{Nature}} \textbf{\bibinfo{volume}{{393}}},
  \bibinfo{pages}{440} (\bibinfo{year}{{1998}}).

\bibitem[{\citenamefont{Albert and Barab\'asi}(2002)}]{albert02}
\bibinfo{author}{\bibfnamefont{R.}~\bibnamefont{Albert}} \bibnamefont{and}
  \bibinfo{author}{\bibfnamefont{A.-L.} \bibnamefont{Barab\'asi}},
  \bibinfo{journal}{Rev. Mod. Phys.} \textbf{\bibinfo{volume}{74}},
  \bibinfo{pages}{47} (\bibinfo{year}{2002}).

\bibitem[{\citenamefont{Barahona and Pecora}(2002)}]{barahona02}
\bibinfo{author}{\bibfnamefont{M.}~\bibnamefont{Barahona}} \bibnamefont{and}
  \bibinfo{author}{\bibfnamefont{L.~M.} \bibnamefont{Pecora}},
  \bibinfo{journal}{Phys. Rev. Lett.} \textbf{\bibinfo{volume}{89}},
  \bibinfo{pages}{054101} (\bibinfo{year}{2002}).

\bibitem[{\citenamefont{Humphries and Gurney}({2008})}]{Humphries08}
\bibinfo{author}{\bibfnamefont{M.~D.} \bibnamefont{Humphries}}
  \bibnamefont{and} \bibinfo{author}{\bibfnamefont{K.}~\bibnamefont{Gurney}},
  \bibinfo{journal}{{Plos One}} \textbf{\bibinfo{volume}{{3}}}
  (\bibinfo{year}{{2008}}).

\bibitem[{\citenamefont{Sen et~al.}(2003)\citenamefont{Sen, Dasgupta,
  Chatterjee, Sreeram, Mukherjee, and Manna}}]{sen03}
\bibinfo{author}{\bibfnamefont{P.}~\bibnamefont{Sen}},
  \bibinfo{author}{\bibfnamefont{S.}~\bibnamefont{Dasgupta}},
  \bibinfo{author}{\bibfnamefont{A.}~\bibnamefont{Chatterjee}},
  \bibinfo{author}{\bibfnamefont{P.~A.} \bibnamefont{Sreeram}},
  \bibinfo{author}{\bibfnamefont{G.}~\bibnamefont{Mukherjee}},
  \bibnamefont{and} \bibinfo{author}{\bibfnamefont{S.~S.} \bibnamefont{Manna}},
  \bibinfo{journal}{Phys. Rev. E} \textbf{\bibinfo{volume}{67}},
  \bibinfo{pages}{036106} (\bibinfo{year}{2003}).

\bibitem[{\citenamefont{Mason and Verwoerd}({2007})}]{Mason07}
\bibinfo{author}{\bibfnamefont{O.}~\bibnamefont{Mason}} \bibnamefont{and}
  \bibinfo{author}{\bibfnamefont{M.}~\bibnamefont{Verwoerd}},
  \bibinfo{journal}{{IET Systems Biology}} \textbf{\bibinfo{volume}{{1}}},
  \bibinfo{pages}{89} (\bibinfo{year}{{2007}}).

\bibitem[{\citenamefont{Boccaletti et~al.}({2006})\citenamefont{Boccaletti,
  Latora, Moreno, Chavez, and Hwang}}]{Boccaletti06}
\bibinfo{author}{\bibfnamefont{S.}~\bibnamefont{Boccaletti}},
  \bibinfo{author}{\bibfnamefont{V.}~\bibnamefont{Latora}},
  \bibinfo{author}{\bibfnamefont{Y.}~\bibnamefont{Moreno}},
  \bibinfo{author}{\bibfnamefont{M.}~\bibnamefont{Chavez}}, \bibnamefont{and}
  \bibinfo{author}{\bibfnamefont{D.~U.} \bibnamefont{Hwang}},
  \bibinfo{journal}{{Physics Reports-Review Section of Physics Letters}}
  \textbf{\bibinfo{volume}{{424}}}, \bibinfo{pages}{175}
  (\bibinfo{year}{{2006}}).

\bibitem[{\citenamefont{Bassett and Bullmore}({2006})}]{bassett06}
\bibinfo{author}{\bibfnamefont{D.~S.} \bibnamefont{Bassett}} \bibnamefont{and}
  \bibinfo{author}{\bibfnamefont{E.}~\bibnamefont{Bullmore}},
  \bibinfo{journal}{{Neuroscientist}} \textbf{\bibinfo{volume}{{12}}},
  \bibinfo{pages}{512} (\bibinfo{year}{{2006}}).

\bibitem[{\citenamefont{Smit et~al.}({2008})\citenamefont{Smit, Stam, Posthuma,
  Boomsma, and de~Geus}}]{smit08}
\bibinfo{author}{\bibfnamefont{D.~J.~A.} \bibnamefont{Smit}},
  \bibinfo{author}{\bibfnamefont{C.~J.} \bibnamefont{Stam}},
  \bibinfo{author}{\bibfnamefont{D.}~\bibnamefont{Posthuma}},
  \bibinfo{author}{\bibfnamefont{D.~I.} \bibnamefont{Boomsma}},
  \bibnamefont{and} \bibinfo{author}{\bibfnamefont{E.~J.~C.}
  \bibnamefont{de~Geus}}, \bibinfo{journal}{{Human Brain Mapping}}
  \textbf{\bibinfo{volume}{{29}}}, \bibinfo{pages}{1368}
  (\bibinfo{year}{{2008}}).

\bibitem[{\citenamefont{Atilgan et~al.}({2004})\citenamefont{Atilgan, Akan, and
  Baysal}}]{Atilgan04}
\bibinfo{author}{\bibfnamefont{A.}~\bibnamefont{Atilgan}},
  \bibinfo{author}{\bibfnamefont{P.}~\bibnamefont{Akan}}, \bibnamefont{and}
  \bibinfo{author}{\bibfnamefont{C.}~\bibnamefont{Baysal}},
  \bibinfo{journal}{{Biophysical Journal}} \textbf{\bibinfo{volume}{{86}}},
  \bibinfo{pages}{85} (\bibinfo{year}{{2004}}).

\bibitem[{\citenamefont{Bartoli et~al.}({2007})\citenamefont{Bartoli,
  Fariselli, and Casadio}}]{Bartoli07}
\bibinfo{author}{\bibfnamefont{L.}~\bibnamefont{Bartoli}},
  \bibinfo{author}{\bibfnamefont{P.}~\bibnamefont{Fariselli}},
  \bibnamefont{and} \bibinfo{author}{\bibfnamefont{R.}~\bibnamefont{Casadio}},
  \bibinfo{journal}{{Physical Biology}} \textbf{\bibinfo{volume}{{4}}},
  \bibinfo{pages}{L1} (\bibinfo{year}{{2007}}).

\bibitem[{\citenamefont{Jespersen and Blumen}(2000)}]{jespersen00}
\bibinfo{author}{\bibfnamefont{S.}~\bibnamefont{Jespersen}} \bibnamefont{and}
  \bibinfo{author}{\bibfnamefont{A.}~\bibnamefont{Blumen}},
  \bibinfo{journal}{Phys. Rev. E} \textbf{\bibinfo{volume}{62}},
  \bibinfo{pages}{6270} (\bibinfo{year}{2000}).

\bibitem[{\citenamefont{Moukarzel and Argollo~de Menezes}(2002)}]{moukarzel02}
\bibinfo{author}{\bibfnamefont{C.~F.} \bibnamefont{Moukarzel}}
  \bibnamefont{and} \bibinfo{author}{\bibfnamefont{M.}~\bibnamefont{Argollo~de
  Menezes}}, \bibinfo{journal}{Phys. Rev. E} \textbf{\bibinfo{volume}{65}},
  \bibinfo{pages}{056709} (\bibinfo{year}{2002}).

\bibitem[{\citenamefont{Sen and Chakrabarti}(2001)}]{sen01}
\bibinfo{author}{\bibfnamefont{P.}~\bibnamefont{Sen}} \bibnamefont{and}
  \bibinfo{author}{\bibfnamefont{B.~K.} \bibnamefont{Chakrabarti}},
  \bibinfo{journal}{Journal of Physics A: Mathematical and General}
  \textbf{\bibinfo{volume}{34}}, \bibinfo{pages}{7749} (\bibinfo{year}{2001}).

\bibitem[{\citenamefont{Moukarzel}(1999)}]{Moukarzel99}
\bibinfo{author}{\bibfnamefont{C.~F.} \bibnamefont{Moukarzel}},
  \bibinfo{journal}{Phys. Rev. E} \textbf{\bibinfo{volume}{60}},
  \bibinfo{pages}{R6263} (\bibinfo{year}{1999}).

\bibitem[{\citenamefont{Kulkarni et~al.}(2000)\citenamefont{Kulkarni, Almaas,
  and Stroud}}]{kulkarni00}
\bibinfo{author}{\bibfnamefont{R.~V.} \bibnamefont{Kulkarni}},
  \bibinfo{author}{\bibfnamefont{E.}~\bibnamefont{Almaas}}, \bibnamefont{and}
  \bibinfo{author}{\bibfnamefont{D.}~\bibnamefont{Stroud}},
  \bibinfo{journal}{Phys. Rev. E} \textbf{\bibinfo{volume}{61}},
  \bibinfo{pages}{4268} (\bibinfo{year}{2000}).

\bibitem[{\citenamefont{Newman and Watts}({1999})}]{newman99}
\bibinfo{author}{\bibfnamefont{M.}~\bibnamefont{Newman}} \bibnamefont{and}
  \bibinfo{author}{\bibfnamefont{D.}~\bibnamefont{Watts}},
  \bibinfo{journal}{{Physics Letters A}} \textbf{\bibinfo{volume}{{263}}},
  \bibinfo{pages}{341} (\bibinfo{year}{{1999}}).

\bibitem[{\citenamefont{Newman and Watts}(1999)}]{newman99_2}
\bibinfo{author}{\bibfnamefont{M.~E.~J.} \bibnamefont{Newman}}
  \bibnamefont{and} \bibinfo{author}{\bibfnamefont{D.~J.} \bibnamefont{Watts}},
  \bibinfo{journal}{Phys. Rev. E} \textbf{\bibinfo{volume}{60}},
  \bibinfo{pages}{7332} (\bibinfo{year}{1999}).

\bibitem[{\citenamefont{Newman et~al.}(2000)\citenamefont{Newman, Moore, and
  Watts}}]{newman00}
\bibinfo{author}{\bibfnamefont{M.~E.~J.} \bibnamefont{Newman}},
  \bibinfo{author}{\bibfnamefont{C.}~\bibnamefont{Moore}}, \bibnamefont{and}
  \bibinfo{author}{\bibfnamefont{D.~J.} \bibnamefont{Watts}},
  \bibinfo{journal}{Phys. Rev. Lett.} \textbf{\bibinfo{volume}{84}},
  \bibinfo{pages}{3201} (\bibinfo{year}{2000}).

\bibitem[{\citenamefont{Almaas et~al.}(2002)\citenamefont{Almaas, Kulkarni, and
  Stroud}}]{almaas02}
\bibinfo{author}{\bibfnamefont{E.}~\bibnamefont{Almaas}},
  \bibinfo{author}{\bibfnamefont{R.~V.} \bibnamefont{Kulkarni}},
  \bibnamefont{and} \bibinfo{author}{\bibfnamefont{D.}~\bibnamefont{Stroud}},
  \bibinfo{journal}{Phys. Rev. Lett.} \textbf{\bibinfo{volume}{88}},
  \bibinfo{pages}{098101} (\bibinfo{year}{2002}).

\end{thebibliography}

\end{document}